\documentclass[sn-apa]{sn-jnl}

\usepackage{graphicx}%
\usepackage{multirow}%
\usepackage{amsmath,amssymb,amsfonts}%
\usepackage{amsthm}%
\usepackage{mathrsfs}%
\usepackage[title]{appendix}%
\usepackage{xcolor}%
\usepackage{textcomp}%
\usepackage{manyfoot}%
\usepackage{booktabs}%
\usepackage{algorithm}%
\usepackage{algorithmicx}%
\usepackage{algpseudocode}%
\usepackage{listings}%
\usepackage{appendix}
\usepackage[T1]{fontenc}
\usepackage{pifont}
\usepackage{siunitx}
\usepackage{tcolorbox}
\usepackage{subcaption}
\usepackage{caption}

\theoremstyle{thmstyleone}%
\theoremstyle{thmstyletwo}%

\theoremstyle{thmstylethree}%

\newif\ifarxiv
\arxivtrue

\newdimen\Cvs \Cvs=3pt
\newdimen\Cdp \Cdp=0pt

  \makeatletter
  \renewcommand{\paragraph}{\@startsection{paragraph}{4}{\z@}%
    {1.0\Cvs \@plus.5\Cdp \@minus.2\Cdp}%
    {.1\Cvs \@plus.3\Cdp}%
    {\reset@font\sffamily\normalsize}
  }
  \makeatother
\begin{document}

\title{Right Move, Right Time: Multi-Sport Space Evaluation Platform for Ultimate Frisbee, Basketball, and Soccer}

\ifarxiv
\author[1]{\fnm{Shunsuke} \sur{Iwashita}}
\author[2,1]{\fnm{Titouan} \sur{Jeannot}}
\author[3]{\fnm{Braden} \sur{Eberhard}}
\author[4]{\fnm{Jacob} \sur{Miller}}
\author[5,1]{\fnm{Rikako} \sur{Kono}}
\author[1]{\fnm{Calvin} \sur{Yeung}}
\author*[1,6]{\fnm{Keisuke} \sur{Fujii}}\email{fujii@i.nagoya-u.ac.jp}

\affil[1]{\orgname{Nagoya University}, \orgaddress{\country{Japan}}}
\affil[2]{\orgname{Aix-Marseille Université}, \orgaddress{\country{France}}}
\affil[3]{\orgname{University of Pennsylvania}, \orgaddress{\country{United States}}}
\affil[4]{\orgname{Shown Space}, \orgaddress{\country{United States}}}
\affil[5]{\orgname{Australian National
University}, \orgaddress{\country{Australia}}}
\affil[6]{\orgdiv{Center for Advanced Intelligence Project}, \orgname{RIKEN}, \orgaddress{\country{Japan}}}


\abstract{
We present an open, sport-agnostic platform that turns tracking into comparable spatial measures across professional Ultimate, basketball, and soccer. Coaches in all three sports ask the same question: where is the usable space, and when should an off-ball run start? Our workflow standardizes inputs, provides timing-aware spatial evaluations, and makes it possible to reuse the same analysis across sports. We illustrate the approach with Ultimate as a focused testbed and then examine transfer between basketball and soccer. Together, these results show a practical path toward consistent, comparable evaluation across various invasion sports.
}
\fi
\maketitle

\section{Introduction}
In many possession-based invasion sports, including NBA basketball, NFL football, MLS soccer, and UFA Ultimate, teams succeed by creating space and acting at the right moment. Yet quantifying that intuition in a way coaches trust is difficult. Coaches and analysts in Ultimate Frisbee (sometimes we call “Ultimate”), basketball, and soccer face the same film-room question: should the receiver without the disc/ball initiate now or wait a beat so that help defense cannot recover? In solving this problem, analysts typically use video and, in some cases, tracking data, yet practical answers remain fragmented by sport, with bespoke workflows that resist reuse and comparison (e.g., \citet{Spearman18,kono2024mathematical,iwashita2025evaluating}). This paper addresses that gap by proposing a sport-agnostic way to evaluate space and timing that coaches can apply across sports, demonstrated through examples in professional Ultimate, basketball, and soccer. Replacing sport-specific silos with common representations and timing-aware measures allows the framework to support reusable workflows, comparable validation, and decisions coaches can trust. 

Our target problem is to establish a sport-agnostic basis for answering the same coaching question across Ultimate Frisbee, basketball, and soccer: where actionable space exists and when an off-ball run should start. To ground the analysis across diverse competitive settings, we draw on NBA basketball and soccer for their mature tracking infrastructures and on the UFA for its emphasis on timing created by the no-dribbling rule. Ultimate Frisbee further extends the framework to emerging invasion sports such as handball and lacrosse. Three concrete challenges make this nontrivial. First, heterogeneity: field geometry, event vocabularies, and sampling rates differ by sport, and many influential models are sport-specific \citep{Spearman18,kono2024mathematical,iwashita2025evaluating}, which hinders apples-to-apples outputs. Second, timing: initiation onsets are unlabeled and embedded in dense trajectories; detecting the start of a cut and linking it to motion constraints and a value signal is difficult even with tracking, and published timing analyses to date are limited to Ultimate amateur practice games \citep{iwashita2025evaluating}. Third, portability and trust: models calibrated in one sport often do not transfer without retuning, and there is no standard for validating cross-sport outputs against observed outcomes and expert expectations. Closing these gaps would allow one workflow to address similar coaching questions across multiple sports.

\begin{figure}[h]
    \centering
    \includegraphics[width=1\linewidth]{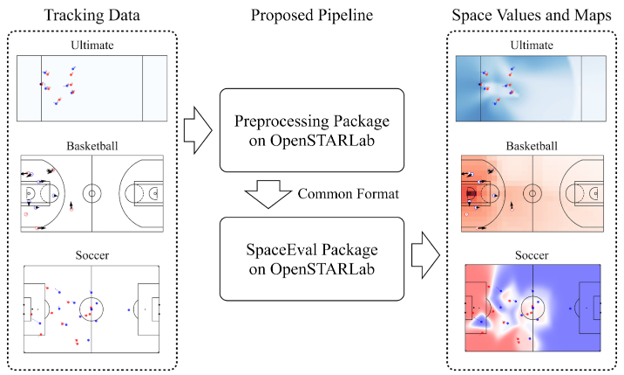}
    \caption{Overview of our multi-sport pitch control platform. Left: The inputs are Ultimate Frisbee, basketball, and soccer tracking data. Middle: sport-agnostic preprocessing and space evaluation via Preprocessing and SpaceEval packages in the open-source Python library called OpenSTARLab. Right: Obtained Ultimate, basketball, and soccer space value maps. 
    Preprocessing Package: \url{https://github.com/open-starlab/PreProcessing} and SpaceEval Package: \url{https://github.com/open-starlab/spaceEval}. 
    }
    \label{fig:diagram}
\end{figure}

In this paper, we introduce an open, pip-installable multi-sport platform that standardizes professional tracking data into a common format and produces sport-agnostic spatial evaluations with sport-specific options, which supports reuse across Ultimate, basketball, and soccer. The goal is to let one worker answer the same question in such sports: where the usable space is and when an off-ball run should start. We include preprocessing utilities and a new module of space evaluation on OpenSTARLab platform \citep{yeung2025openstarlab} for UFA, NBA, and World Cup datasets. Figure~\ref{fig:diagram} summarizes the pipeline and representative outputs across sports. We then apply the workflow to Ultimate Frisbee as a focused testbed and examine transfer between basketball and soccer under the same evaluation scheme. 

\begin{figure}[h]
    \centering
    \includegraphics[width=1\linewidth]{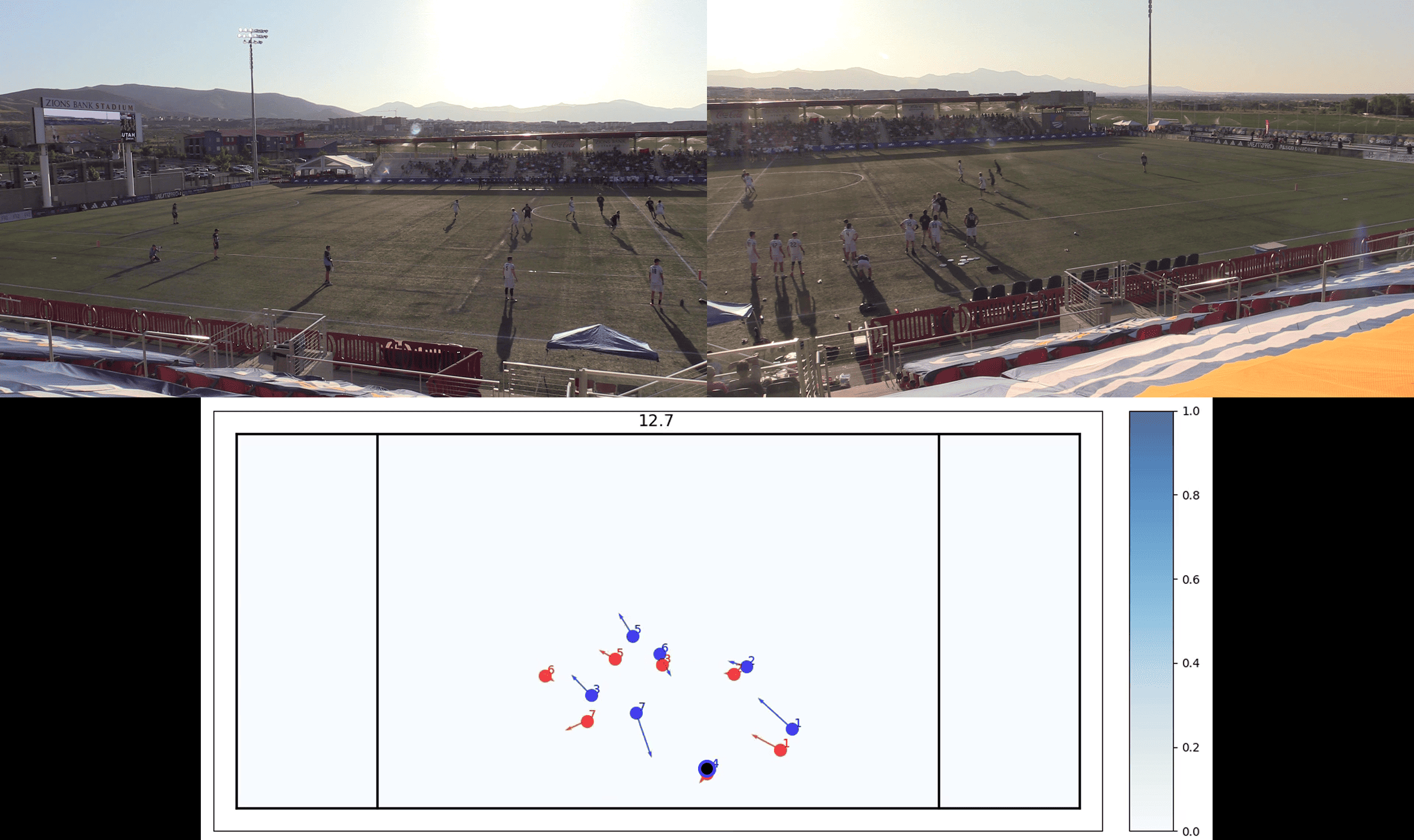}
    \caption{UFATrack dataset. This dataset is derived from a part of the official UFA match footage between the Oakland Spiders and the Salt Lake Shred held on June 27, 2025. The data contain detailed spatio-temporal tracking annotations of player xy coordinates. This dataset anchors the Ultimate analyses and is released for reproducible research. The dataset is shared at: \url{https://zenodo.org/records/17840132}}
    \label{fig:ufatrack}
\end{figure}

Our contributions are threefold. First, we provide a general framework and workflow that makes multi-sport pitch control analysis practical within one platform. It standardizes inputs, introduces timing-aware spatial measures, and exposes sport-specific options explicitly. Second, we release a professional Ultimate Frisbee dataset (called UFATrack, as illustrated in Figure~\ref{fig:ufatrack}) with synchronized video, tracking, and event annotations to support reproducible studies and practitioner use. Using this dataset, timing-aware analysis of off-ball run initiation is demonstrated for Ultimate using a counterfactual approach \citep{iwashita2025evaluating}. Third, we assess cross-sport transfer by adapting a basketball model that accounts for dribbling and interception, BIMOS (Ball Intercept and Movement for Off-ball Scoring)\citep{kono2024mathematical}, to soccer and evaluating it on World-Cup 2022 shot scenes against a soccer baseline, OBSO (off-ball scoring opportunities)\citep{Spearman18}, finding that the adapted model aligns more closely with observed outcomes.

The remainder of this paper is as follows.
\textbf{Section~2} explains the background of this work, and \textbf{Section~3} details preprocessing and space evaluation pipelines, including pitch-control families, the space and play evaluation metrics using the datasets spanning Ultimate, basketball, and soccer. \textbf{Section~4} describes the UFATrack and WC2022 (FIFA World Cup 2022) datasets we used. \textbf{Section~5} reports validation of the models in UFA datasets and transfer results from the basketball model to WC2022 datasets. Finally, in \textbf{Section~6}, we applied the platform to a UFA match case study (SLC vs. OAK) to show how initiation timing affects the game results, and \textbf{Section~7} concludes this paper.

\section{Background}
\subsection{Related work}
On-ball player performance is often modeled via event prediction, while off-ball movement is commonly assessed through spatial metrics. Representative mathematical approaches include dominant regions based on Voronoi geometry and minimum-arrival-time ideas \citep{taki1996development, fujimura2005geometric}, later extended toward player-specific kinematics and space weighting \citep{brefeld2019probabilistic, martens2021space, narizuka2021space}. More recent probabilistic formulations estimate OBSO \citep{Spearman18} and quantify space-creating movements among attackers and defenders \citep{Fernandez18}. 

Beyond these core threads, rule-based evaluations of off-ball actions in basketball examine cutting and its impact on open shots \citep{supola2023modeling, lamas2015modeling}. In soccer, rule-based notions of dangerousness have been explored \citep{link2016real}. Building on OBSO \citep{Spearman18}, researchers extended scoring with defender modeling and other variants \citep{teranishi2022evaluation, yeung2024strategic, ogawa2025pitch}, adapted ideas to defenders in soccer \citep{umemoto2023evaluation}, to attackers in basketball \citep{kono2024mathematical}, and to Ultimate \citep{iwashita2025evaluating}. In particular, \citet{kono2024mathematical} proposed BIMOS, which models pass-to-score, dribble-to-score, and interception contexts. 

\subsection{Motivation to Focus on Ultimate Frisbee}
Ultimate Frisbee serves as the starting point for this study, giving us a useful focal sport that anchors our cross-sport analysis. Ultimate shares core features with basketball, American football, and soccer, including coordinated possessions, spatial pressure, and timing-dependent openings, while offering analytically useful constraints (no dribbling, a stationary thrower, a fixed stall count, and largely continuous play) that make off-ball initiation timing more observable and consequential. Using it as a testbed helps establish methods that can be transferred to other developing or rapidly professionalizing sports, where data pipelines and analytic norms are still taking shape. Sports such as lacrosse, rugby sevens, team handball, and even newer formats like small-sided soccer or mixed-gender field sports face similar questions about space creation and off-ball decision making. Demonstrating robust, portable analytics in Ultimate can set a template for these emerging domains, helping them adopt consistent evaluation frameworks early in their development.

Our analysis follows the men’s professional Ultimate Frisbee Association (UFA) format. Two teams of seven play on a field measuring 80 yards by 53 and one-third yards, with 20-yard end zones. Players may pass in any direction, must set a pivot foot after a catch, and the thrower must release within a 7-second stall count tracked by officials. Games consist of four 12-minute quarters, and play is largely continuous; outside of goals, fouls, or timeouts, the flow rarely stops. Possession changes on incompletions, out-of-bounds throws, interceptions, or stalls, and the defense takes over at the spot to attack the opposite end zone. A completed pass into the end zone scores one point, after which the scoring team executes a “pull”, a long restart throw similar to a kickoff.

Work on Ultimate analytics is limited but growing. Early studies introduced location-based completion and scoring probability maps to quantify the value of field regions, which were then extended to a player-contribution metric \citep{weiss2013maps, weiss2014spatial}. Passes and turnovers have been modeled as zone-to-zone state transitions \citep{lam2021state}, while league-facing impact summaries have also appeared, such as an AUDL Player Efficiency Rating \citep{geertsen2021}. Using professional tracking, \citep{eberhard2025machine} combined four seasons of data to train completion-probability and field-value models that jointly reflect throw difficulty and positional value. Most prior Ultimate work relies on discrete event logs rather than continuous trajectories, which constrains off-ball initiation analysis; even with tracking, two hurdles persist: detecting the onset of a run within dense trajectories and linking that moment to a motion constraint and value signal with modest samples. For Ultimate specifically, pitch-control-based space valuation has been adapted from soccer using drone video, and a three-on-three dataset applied a modified spatial metric without addressing timing \citep{iwashita2024space}. We named Counterfactual-Ready Space Value (CRSV), which measures the frame-level value of space derived from reachability-based pitch control and is defined so that shifting the initiation time of an off-ball run yields valid counterfactual comparisons. Recent work directly evaluates initiation timing via temporal counterfactuals in Ultimate \citep{iwashita2025evaluating}. Since Ultimate prohibits dribbling, requires a pivot, and enforces a stall count while sharing invasion-sport structure with basketball and soccer, it offers a clean, comparable testbed for studying when to start an off-ball run before generalizing to other sports.

\section{Methods}
This section introduces the complete pipeline as follows.
\begin{itemize}
    \item 3.1. Preprocessing Package for Space Data Format, which converts various tracking and event logs into a unified space-data representation ready for analysis;
    \item 3.2. SpaceEval Package for Multi-Sport Pitch Control, which provides a common interface to compute, compare, and visualize pitch-control and space-value models;
    \item 3.3. OBSO framework, which specifies the basis of reachability-based control and off-ball scoring opportunity formulations;
    \item 3.4. CRSV framework, which refines control fields and defines frame-level spatial value for downstream analyses;
    \item 3.5. BIMOS framework, which incorporates ball delivery and interception to evaluate scoring opportunity creation; and
    \item 3.6. Evaluation Metrics, which describes the verification methodologies.
\end{itemize}

\subsection{Preprocessing Package for Space Data Format}
The \textbf{Space Data Module} of the openstarlab-preprocessing package \citep{yeung2025openstarlab} adds preprocessing routines for soccer and Ultimate that convert raw tracking or positional data into a structured and standardized space-based representation. A single generalized Space data class provides the shared structure for loading match metadata, organizing team and player information, storing pitch dimensions, and managing frame-by-frame positional data across both soccer and Ultimate. The corresponding preprocessing pipeline handles tasks such as merging raw files, sorting by timestamp, aligning coordinate systems, cleaning missing values, attaching team and period labels, identifying on-field players, and constructing unified per-frame data structures used for downstream spatial and event-based analysis. This module provides a consistent and standardized input format across sports, enabling direct integration with the broader OpenSTARLab preprocessing and modeling pipelines.

\subsubsection{Key features}
The key features of the openstarlab-preprocessing package Space Data Module are as follows:
\begin{itemize}
    \item \textbf{Unified Space Class:}  
    The module provides a unified Space class that organizes all core match information, including metadata, team and player identities, pitch dimensions, and frame-by-frame positional data. This centralized structure ensures that the soccer, basketball, and Ultimate data have a consistent representation, simplifying interaction with downstream components.

    \item \textbf{Standardized Pipelines:}  
    The preprocessing pipelines for soccer and Ultimate merge raw input files, sort and align timestamps, and normalize coordinate fields to produce coherent spatio-temporal sequences. These routines also handle basic cleaning and the matching of event and tracking data where applicable, ensuring robustness and consistency across different data sources.

    \item \textbf{Contextual Feature Generation:}  
    The module enriches each frame with essential contextual information such as team labels, period indices, and other match-level metadata. These derived fields provide temporal and organizational structure to the positional data, ensuring that each coordinate is situated within the correct team and match context for downstream spatial analysis.

    \item \textbf{Analysis-Ready Output:}  
    The final output is a standardized space-data format designed for seamless integration with the broader OpenSTARLab pipeline. This representation is clean, structured, and compatible with SpaceEval and modeling modules, enabling efficient transition from raw data to analysis or model development.
\end{itemize}

\subsubsection{Space data format}
The preprocessing package produces a sport-specific data format that retains the characteristics of each dataset while enforcing clean formatting, consistent temporal alignment, and reliable coordinate handling. For soccer, event logs and optical-tracking streams are reshaped into structured DataFrames that preserve the full sequence of match actions and frame-by-frame positional information, summarized in Tables~\ref{tab:event_dataframe} and~\ref{tab:tracking_dataframe}, respectively. For Ultimate, the structure is similar to soccer (with a different number of players and a disc instead of a ball). These processed outputs ensure that events, player trajectories, and ball movements can be linked without ambiguity, supporting downstream tasks such as pitch-control computation and spatial value modeling.

\begin{table}[h!]
\centering
\caption{Description of the variables in the event data after preprocessing, which is common among sports.}
\label{tab:event_dataframe}
\begin{tabular}{lll}
\toprule
\textbf{Variables} & \textbf{Type} & \textbf{Description} \\
\midrule
Team & str & Team associated with the event ('Home' or 'Away') \\
Type & str & Type of the event \\
Subtype & str & Subtype of the event (e.g., success, fail) \\
Period & int & Match period (1 or 2) \\
Start Frame & int & Frame index at event start \\
Start Time [s] & float & Time in seconds at event start \\
End Frame & int & Frame index at event end \\
End Time [s] & float & Time in seconds at event end \\
From & str & Player performing the action \\
To & str & Player receiving the action (if applicable) \\
Start X & float & Starting x-coordinate of the event \\
Start Y & float & Starting y-coordinate of the event \\
End X & float & Ending x-coordinate of the event \\
End Y & float & Ending y-coordinate of the event \\
\bottomrule
\end{tabular}
\end{table}

\begin{table}[h!]
\centering
\caption{Description of the variables in the dataset home and away tracking data after preprocessing. The number K means the total number of recorded players, which depends on the sport.}
\label{tab:tracking_dataframe}
\begin{tabular}{lll}
\toprule
\textbf{Variables} & \textbf{Type} & \textbf{Description} \\
\midrule
Period & int & Match period (1 or 2) \\
Time [s] & float & Time in seconds \\
Home\_1\_x ... Home\_K\_x & float & X-coordinates of home players 1--K \\
Home\_1\_y ... Home\_K\_y & float & Y-coordinates of home players 1--K \\
Away\_1\_x ... Away\_K\_x & float & X-coordinates of away players 1--K \\
Away\_1\_y ... Away\_K\_y & float & Y-coordinates of away players 1--K \\
ball\_x & float & Ball X-coordinate \\
ball\_y & float & Ball Y-coordinate \\
\bottomrule
\end{tabular}
\end{table}

\subsubsection{Customization and flexibility}
The openstarlab-preprocessing package provides configurable preprocessing options, allowing users to tailor the pipeline to different project needs. Its modular design supports seamless integration into larger data-processing workflows. For additional information, please refer to the OpenSTARLab documentation\footnote{Comprehensive dataset documentation is available at: \url{https://openstarlab.readthedocs.io/en/latest/Pre_Processing/Sports/Space_Eval/Data_Provider/index.html}}. The example below illustrates how to preprocess the World Cup soccer dataset using the openstarlab-preprocessing package. For UFA Ultimate and NBA Basketball, we can apply it with similar codes.

\begin{tcolorbox}[colback=gray!5!white, colframe=gray!75!black, title=Python code example for preprocessing World Cup Soccer dataset] 
\begin{verbatim} 
from preprocessing import Space_data 

Space_data(data_provider="fifa_wc_2022", 
            event_data_path="path/to/event/folder", 
            tracking_data_path="path/to/tracking/folder", 
            out_path="path/to/save/preprocessed/files" 
            ).preprocessing() 
\end{verbatim} 
\end{tcolorbox}

In summary, the Space Data Module offers a robust and efficient solution for pre-processing spatial event data (Table~\ref{tab:event_dataframe}) and tracking data (Table~\ref{tab:tracking_dataframe}) across sports, converting heterogeneous raw inputs into a unified and standardized space-data format. This streamlined representation enables consistent downstream analysis and supports seamless integration with the broader OpenSTARLab evaluation and modeling frameworks.

\subsection{SpaceEval Package for Multi-Sport Pitch Control}
\subsubsection{Overview of SpaceEval Package}
The SpaceEval package provides a unified framework for computing, comparing, and benchmarking spatial evaluation models across multiple sports. Building on standardized space-data inputs, it offers tools for calculating pitch-control, space value, and other spatial metrics, as well as interfaces for implementing or adapting sport-specific and cross-sport models. The package supports analyses ranging from off-ball movement evaluation to counterfactual assessments of player decisions, and enables direct comparison of models originally developed for different sports. By consolidating evaluation pipelines and providing common metrics, SpaceEval facilitates reproducible research, cross-sport transfer studies, and systematic evaluation of spatial behavior in soccer, basketball, and Ultimate.

The key features of the SpaceEval package are as follows:
\begin{enumerate}
    \item \textbf{Unified Space Evaluation Class}:  
    The package provides a unified class for running space-evaluation models across different sports. Through the SpaceModel class, users can specify event and tracking inputs, configure model options, and execute full-space evaluation workflows. This design ensures a consistent user experience regardless of whether the analysis targets soccer or Ultimate.

    \item \textbf{Sport-Specific Space Models}:  
    SpaceEval includes sport-aware model implementations tailored to the requirements of each domain. For soccer, the package provides the OBSO \citep{Spearman18} and BIMOS \citep{kono2024mathematical} models, enabling the computation of pitch control and shooting opportunity metrics. For Ultimate, the module adapts pitch-control logic to Ultimate-specific rules such as stationary throwers and disc movement constraints, allowing for evaluation of receiving opportunities and space creation.

    \item \textbf{Batch Processing and Dataset-Level Evaluation}: 
    The system supports dataset-level evaluation by allowing users to process multiple matches or scenes in batch mode. Given a directory of preprocessed space-data files, SpaceEval automatically iterates across all samples, computes spatial metrics, and returns structured outputs indexed by match or event. This facilitates large-scale model evaluation and comparison.

    \item \textbf{Built-in Visualization Tools}:  
    The package offers integrated visualization utilities for rendering model outputs directly onto standardized soccer and Ultimate fields (for example, see Section~\ref{sec:result}). These include heatmaps for pitch control, overlays of attacking/defensive areas, and event-level scene visualizations for model debugging or qualitative analysis. The visual outputs support both research communication and practitioner interpretation.
\end{enumerate}

The example below illustrates how to analyze the UFA dataset using the openstarlab-spaceEval package.

\begin{tcolorbox}[colback=gray!5!white, colframe=gray!75!black, title=Python code example for evaluating space in UFA Ultimate dataset] 
\begin{verbatim} 
from spaceeval import Space_Model

Space_Model(space_model="wUPPCF",
            event_data="path/to/event/folder",
            tracking_home="path/to/tracking_home/folder",
            tracking_away="path/to/tracking_away/folder",
            provider="UFA",
            out_path="path/to/save/spaceeval/files",
            )
\end{verbatim} 
\end{tcolorbox}
Next, we introduce OBSO, CRSV, and BIMOS frameworks implemented by the openstarlab-spaceEval package.

\subsubsection{OBSO Framework \citep{Spearman18}}
The off-ball scoring opportunity (OBSO) quantifies the likelihood that the attacking team will score with the next on-ball action, including contributions from off-ball players who do not currently possess the ball. The OBSO is formulated as a decomposition over all possible pitch locations $\mathbf{r} \in \mathbb{R}^2$:
\begin{equation}
    P(G \mid D)
        = \sum_{\mathbf{r} \in \mathbb{R}^2}
            P(G_\mathbf{r} \mid C_\mathbf{r}, T_\mathbf{r}, D)\,
            P(C_\mathbf{r} \mid T_\mathbf{r}, D)\,
            P(T_\mathbf{r} \mid D),
    \label{eq:obso}
\end{equation}
where
\begin{itemize}
    \item $D$ represents the instantaneous game state (player positions, velocities, and contextual information),
    \item $T_r$ denotes the probability that the next on-ball event occurs at location $\mathbf{r}$,
    \item $C_r$ denotes the probability that the attacking team successfully controls the ball at $\mathbf{r}$, and
    \item $G_r$ denotes the probability of scoring from location $\mathbf{r}$.
\end{itemize}

The first and third terms are score and transition models, respectively. The second term, $P(C_r \mid T_r, D)$, is modeled using the \textit{Potential Pitch Control Field} (PPCF), which describes the probability that each player  can reach and control the ball at location  within a given time horizon. The model assumes that, when a player is in proximity to the ball, their ability to make a controlled touch can be represented as a Poisson point process. In this formulation, the longer a player remains near the ball without interference from an opponent, the greater the probability that they will successfully gain control. For simplicity, we can assume that $P(G_\mathbf{r}|D)$, $P(T_\mathbf{r}|D)$, $P(C_\mathbf{r}|D)$ are independent (e.g., \citet{teranishi2022evaluation}).

For each player $j$, the temporal evolution of their control probability is given by:
\begin{align}
\frac{d\,\mathrm{PPCF}_j}{dT}(t, \mathbf{r}, T \mid s, \lambda_j)
&=
\Big( 1 - \sum_k \mathrm{PPCF}_k(t, \mathbf{r}, T \mid s, \lambda_j) \Big)\,
    f_j(t, \mathbf{r}, T \mid s)\,
    \lambda_j, \label{eq:ppcf_diff} \\[4pt]
f_j(t, \mathbf{r}, T \mid s)
&=
\bigg[1 + \exp\!\bigg(
    -\pi \frac{T - \tau_{\mathrm{exp}}(t, \mathbf{r})}{\sqrt{3}\,s}
\bigg)\bigg]^{-1}, \label{eq:logistic_intercept}
\end{align}
where:
\begin{itemize}
    \item $f_j(t,\mathbf{r},T\mid s)$ is the probability that player $j$ arrives at location $\mathbf{r}$ within time $T$, modeled using a logistic distribution centered around their expected intercept time,
    \item $\tau_{\mathrm{exp}}(t,\mathbf{r})$ is the expected time for player j to reach $\mathbf{r}$, computed using a kinematic model with player location, acceleration, and maximum sprint speed,
    \item $\lambda_j$ is the player’s control-rate parameter (attackers and defenders may take different values), and
    \item $s$ is a shape parameter capturing uncertainty in player arrival time.
\end{itemize}

Integrating Equation~\ref{eq:ppcf_diff} over all $T \ge 0$ yields each player's marginal control probability at location $\mathbf{r}$, and summing over all attackers provides the final PPCF field:
\begin{equation}
\mathrm{PPCF}(t,r)
    = \sum_{j \in \text{attackers}} \mathrm{PPCF}_j(t,r).
\end{equation}
PPCF is assumed to be zero before the time of flight of the ball, $T_{flight}$, at location $\mathbf{r}$. This field serves as the spatial foundation for both the transition model and the combined OBSO computation described above.

\subsubsection{CRSV framework \citep{iwashita2025evaluating}}
Unlike the OBSO family, which scores off-ball eventual scoring opportunities from locations, CRSV quantifies the usable space values aligned with reachability and game constraints. It is \textbf{counterfactual-ready}: shifting the initiation time of an off-ball run yields valid comparisons by construction. In short, CRSV complements OBSO by focusing on \emph{when} space is actually exploitable, not only \emph{where} it is promising.

\paragraph{Weighted pitch control in Ultimate (wUPPCF)}
Building on \citep{Spearman18} and its Ultimate-specific extension \citep{iwashita2024space} called UPPCF (Ultimate PPCF), we redefine the weighted pitch-control formulation introduced in \citep{iwashita2025evaluating} called wUPPCF (weighted Ultimate PPCF) within the CRSV framework. OBSO \citep{Spearman18} estimates the eventual probability of scoring from a location by chaining pass feasibility and shot value, which makes it an outcome-centric measure. UPPCF \citep{iwashita2024space} and wUPPCF \citep{iwashita2025evaluating} adapt pitch control to Ultimate and quantify who can realistically claim or use space at a given moment under sport-specific constraints such as reaction and throwing limits, which is a moment-to-moment spatial control metric rather than a scoring probability. The field downweights far targets and potential block lanes for receiver $j$ such that:
\begin{equation}
\text{wUPPCF}_j(t,\mathbf{r}) = \text{UPPCF}_j(t,\mathbf{r}) w_d(t,\mathbf{r}) w_s(t,\mathbf{r}).
\end{equation}
Here $w_d$ penalizes distance from the disc (throw difficulty), and $w_s$ accounts for marker obstruction along the intended trajectory. 

\paragraph{Frame- and scenario-level values}
Let $\Omega_j(t)$ denote the predicted \emph{simultaneous-arrival} region where receiver $i$ and the disc can meet under the motion and throw model. We define the frame-wise value as
\begin{equation}
V_{\text{frame}}(t) \;=\; \frac{1}{|\Omega_j(t)|} \sum_{\mathbf{r}\in\Omega_j(t)} \text{wUPPCF}_j(t,\mathbf{r}).
\end{equation}
For a timing-shifted counterfactual scenario indexed by $\xi$ (advance/delay of the initiation), the scenario value is the maximum of a moving average over a short window $w$:
\begin{equation}
V_{\text{scenario}}^{(\xi)} \;=\; \max_{t} \frac{1}{w}\sum_{k=1}^{w} V_{\text{frame}}(t+k).
\end{equation}

\paragraph{Counterfactual motion model}
We generate a family of temporally shifted trajectories by changing only the initiation frame $t_0$ of the target receiver by an offset $\xi$ (earlier or later) while keeping all other players’ trajectories identical to the original play (for detecting movement initiation timing, see the original paper \citep{iwashita2025evaluating}).
This isolates the causal effect of timing on spatial control and yields like-for-like comparisons across scenarios. Concretely, the method constructs counterfactual plays by shifting the start of the receiver’s run forward or backward in time and then evaluating each scenario with the same space metric; only the initiation timing differs across scenarios, not teammates, opponents, or context. This design choice follows the paper’s rule-based approach, emphasizing interpretability and controllability of the timing variable.

For an \emph{earlier} initiation ($\xi<0$), the receiver’s original movement is replayed $|\xi|$ frames earlier and translated so that the trajectory connects continuously at the new initiation point; the correction vector ensures no discontinuity at $t_0+\xi$. For a \emph{delayed} initiation ($\xi>0$), a short bridging segment is inserted first: the receiver moves linearly from the original position at $t_0$ using the average pre-initiation velocity (computed over the preceding one second), after which the original trajectory is followed with a continuity correction. This construction fills the temporal gap introduced by delaying the start and maintains a realistic path while altering only the timing. All other players retain their original trajectories so that the only manipulated factor is $\xi$. 

\paragraph{Timing effectiveness}
Finally, the timing metric compares the realized timing with the best counterfactual:
\begin{equation}
V_{\text{timing}} \;=\; V_{\text{scenario}}^{(0)} - \max_{\xi\neq 0} V_{\text{scenario}}^{(\xi)}.
\end{equation}
A larger $V_{\text{timing}}$ indicates that the chosen initiation timing exploited usable space better than any small shift; smaller values suggest alternative starts may have been preferable.

\textbf{Remarks.} CRSV is model-agnostic with respect to the underlying pitch-control variant (e.g., PPCF, UPPCF, and wUPPCF).

\subsubsection{BIMOS Framework \citep{kono2024mathematical} }
    The BIMOS model was designed to address a limitation of OBSO regarding interception possibilities; OBSO assumes that all players move toward location $r$ rather than selecting alternate paths that may offer a higher chance for ball interception\footnote{This was partially considered in a prior study~\citep{Spearman2017}.}. The transition and control models in Equation~\ref{eq:obso} are combined, and BIMOS is formulated as:
    \begin{equation}
        P(G \mid D)
            = \sum_{\mathbf{r} \in \mathbb{R}^2}
                P(G_\mathbf{r} \mid C_\mathbf{r}, T_\mathbf{r}, D)\,
                P(C_\mathbf{r}\,\cap\,T_\mathbf{r} \mid D).
        \label{eq:BIMOS}
    \end{equation}
    The first term is the score model, which is conceptually the same as OBSO. 
    The second term, $P(C_\mathbf{r}\,\cap\,T_\mathbf{r} \mid D)$, is defined as the \textit{Potential Ball Control Field} (PBCF), which estimates the probability that the attacking team delivers and controls the ball upon sending it to location $\mathbf{r}$. PBCF calculates its value at the dynamically changing “ball location” for each time horizon, whereas PPCF uses a fixed target position, thereby leading to the integration being carried out over $0 \leq T \leq T_{\mathrm{flight}}$, from the start to the end of its delivery, with slight modifications to the definition of $f_j(t,\mathbf{r},T)$ such that:
    \begin{itemize}
        \item $f_j(t,\mathbf{r},T\mid s)$ is the probability that player $j$ arrives at ``ball location'' within time $T$, modeled using a logistic distribution centered around their expected intercept time,
        \item $\tau_{\mathrm{exp}}(t,\mathbf{r})$ is the expected time for player $j$ to reach ``ball location'', computed using a kinematic model with player location, acceleration, and maximum sprint speed.
    \end{itemize}
    BIMOS is further extended to dribbling situations by adjusting parameters such as ball speed and the set of attackers involved in the calculation, and the total BIMOS is then obtained by summing the pass and dribble components with weighting\footnote{The weighting is currently determined based on the observed tendency rates for passes and dribbles. Alternatively, the final BIMOS can be taken as the larger of the two values.}.

\subsection{Evaluation Metrics}
\subsubsection{Ultimate Frisbee}
To examine the validity of the frame-level values, we conducted a team-level comparison of the proposed indicators in an actual game setting. For each sequence, we computed $V_{frame}$ at every frame following the receiver’s initiation and adopted its maximum value as a representative measure of the quality of the attacking opportunity in that play. We then compared the distribution of these values between the two teams.

Next, to evaluate timing optimality, we compute $V_{timing}$ per possession as defined above and aggregate to the team level, and compare its distribution between the teams. Based on this, we evaluated the extent to which these timing-related indicators could account for the match outcome and the final score margin.

Furthermore, to capture within-possession dynamics of the offense controls space during a possession, we form a time series of a high-control ratio at each frame, defined as the share of pitch locations with wUPPCF$\leq\tau$ ($\tau$ is a threshold). The series is averaged every $\Delta$ s (we set $\Delta = 5$ s) and summarized by simple statistics such as the peak value and time-above-$\tau$. We then compare these summaries across outcome classes (for example, turnover vs. goal) to assess whether stronger control precedes positive outcomes.

\subsubsection{Soccer}
To assess the reliability and predictive strength of our scoring-opportunity models, we first compute the Pearson correlation between performance indicators in the current match ($i$-th game)
and the subsequent match ($i+1$-th game). Specifically, we evaluate the correlations of OBSO,
BIMOS, shots, and goals for each team in the World Cup dataset. Unlike the previous study
\citep{Spearman18}, which conducted this analysis at the player level, we adopt a team-level
perspective due to the limited number of matches per team and the fact that many players register
few or no shots, making player-level correlations unstable.

Secondly, the evaluation focused on a prediction task in which the values produced by OBSO and BIMOS were used as the estimated probability that a given shot would result in a goal. These predicted probabilities were then compared against the actual outcomes to assess how well each model captured true scoring likelihood. Probabilistic accuracy was evaluated using the log-loss, which is mathematically equivalent to maximizing the likelihood of the observed outcomes. In this sense, log-loss corresponded directly to the Maximum Likelihood Estimation (MLE) principle. The core idea behind MLE was to identify the model whose predicted probabilities were most consistent with the underlying ground-truth distribution (i.e., whether a shot resulted in a goal). Models that assigned higher probability to the correct outcomes achieved lower log-loss, whereas models that were miscalibrated or overconfident were penalized more heavily.

\section{Datasets}
\subsection{UFATrack dataset}
For the Ultimate domain, we constructed and released the UFATrack dataset for this study (Figure~\ref{fig:ufatrack}). This dataset is derived from the game footage of the second quarter of an official UFA match between the Oakland Spiders and the Salt Lake Shred (score: OAK 4–9 SLC) held on June 27, 2025. We focus on the second quarter because Salt Lake Shred overturned a deficit from the first quarter and built a substantial lead, a situation in which we expected large differences in team performance to emerge.

The second quarter in UFA consists of 12 minutes of game time; in our dataset, we include
approximately 9.5 minutes of effective play, excluding pulls and offensive-defensive transitions. In
total, the data contain 20 possession sequences, with detailed spatio-temporal annotations based
on precise (x, y) coordinates in metric units, defined on a coordinate system corresponding to a
standard pitch of 109.73 m × 48.77 m (equal to 80 yards by 53 and one-third yards with 20-yard
end zones). For each frame, the positions of all 14 on-field players and the disc are provided. The
disc position is obtained by annotating the disc holder and, for frames in which no player is in
possession, linearly interpolating the disc location between consecutive possession events. The
source video is recorded at 30 frames per second (fps), and tracking data are sampled at 10 fps.
This enables high spatial-resolution tracking of both player and disc movements, and is expected to
serve as a benchmark dataset for evaluating pitch-control models and space-utilization metrics in
Ultimate.

\subsection{World Cup Soccer dataset}
For the soccer domain, we utilized the StatsBomb 2022 World Cup Dataset\footnote{StatsBomb 2022 World Cup Dataset: \url{https://blogarchive.statsbomb.com/news/statsbomb-release-free-2022-world-cup-data/}}, a publicly available open dataset released by Hudl StatsBomb. This dataset provides comprehensive event-level and freeze-frame data for all 64 matches of the FIFA World Cup Qatar 2022, encompassing the data of 32 national teams over the tournament. The data contains detailed spatio-temporal annotations of on-ball actions such as passes, carries, shots, dribbles, pressures, duels, and recoveries, with precise (x, y) coordinates on a normalized 100×100 pitch grid. In total, the dataset includes approximately 120,000 events across the 64 matches, averaging around 1,875 events per game, making it one of the most granular public datasets for international soccer. In addition to event streams, the release also includes ``StatsBomb 360'' freeze-frames, which provide the positional context of all 22 players (and the ball) at the instant of key on-ball events. This contextual layer enables frame-level spatial analysis similar to full optical tracking systems, though at a lower temporal density.

To integrate the dataset into our unified SpaceEval pipeline, all event coordinates were converted from StatsBomb’s normalized grid to metric units corresponding to a 105 m × 68 m standard pitch. The dataset’s high spatial precision and detailed event annotations make it well-suited for evaluating pitch-control models and space-value representations in professional soccer. Within our multi-sport framework, the World Cup dataset functions as the reference benchmark for soccer, enabling consistent cross-sport comparisons of spatial dynamics between elite soccer, professional basketball, and Ultimate under a unified evaluation methodology.

\section{Results: Ultimate validation and cross-sport transfer}
\label{sec:result}
This section tests two questions. First, in Ultimate, do our space measures behave coherently across teams and within possessions, and what does a single possession reveal when we vary initiation timing in a controlled way. Second, in soccer, do outcome models for a different sport transfer reliably when judged by temporal stability, calibration, and how they represent control along the ball path.

\subsection{Ultimate: Validation and Case Study}
We evaluate Ultimate at three levels that mirror how analysts read games: (1) team-level distributions from wUPPCF and $V_{frame}$, (2) within-possession dynamics using a high-wUPPCF area ratio versus turnovers and goals, and (3) a single-possession counterfactual that varies the receiver’s start time to illustrate what changes when initiation is earlier or later.

\subsubsection{Team-level signals: wUPPCF \& $V_{frame}$}
We first examine the wUPPCF to check whether frame-level space values capture differences in play quality at the team level and whether timing-only indicators account for outcomes. This analysis serves as an interpretability and face-validity check for CRSV framework: if the metric tracks meaningful offensive opportunity and clarifies what timing adds, it may be suitable for practical use.

\begin{figure}[b]
    \centering
    \includegraphics[width=1\linewidth]{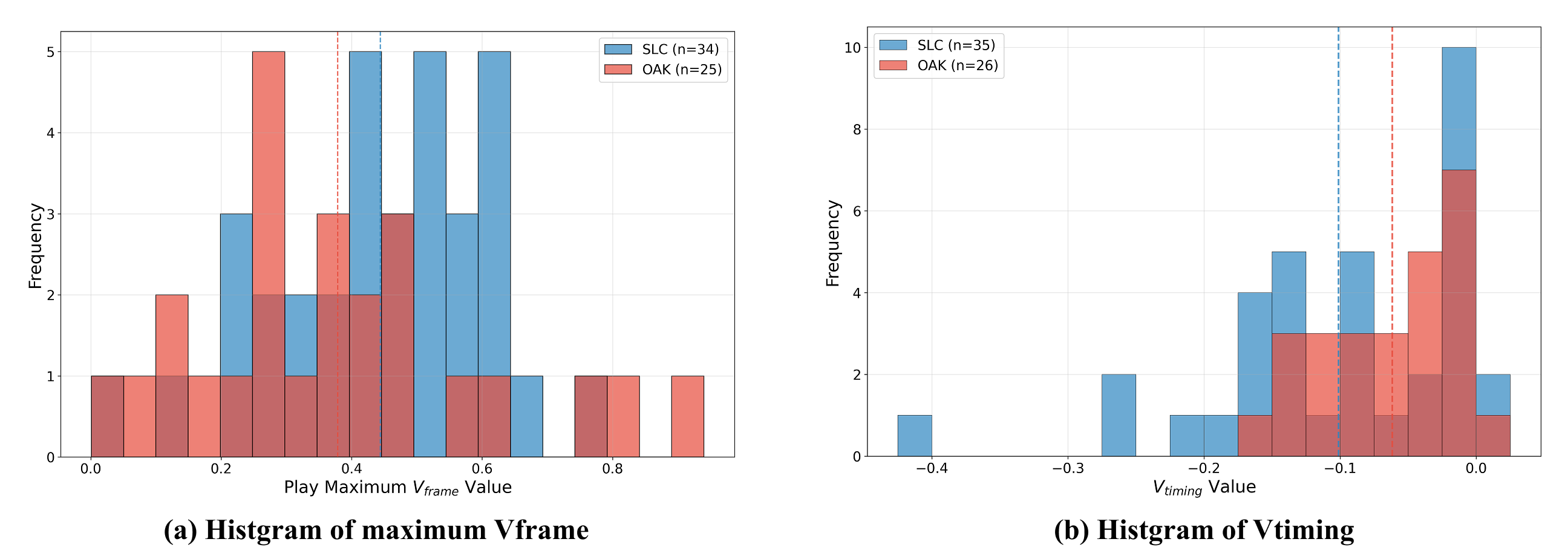}
    \caption{(a) Histogram of the maximum $V_{frame}$, showing that SLC generated a larger number of medium-quality plays, whereas OAK produced fewer but higher-quality plays. (b) Histogram of $V_{timing}$, which did not differ substantially between the teams, indicating that the five-point margin may not reflect a true difference in overall outcome.}
    \label{fig:wuppcf_results}
\end{figure}

When comparing the performances of the two teams, SLC (who led 9–4 in the second quarter) exhibited a distribution based on the maximum $V_{frame}$ in each sequence in which medium-quality plays accounted for a large proportion (blue in Figure~\ref{fig:wuppcf_results}(a)), whereas OAK, despite having fewer attempts overall, generated a certain number of high-quality plays, while low-quality plays made up the majority of its distribution (red in Figure~\ref{fig:wuppcf_results}(a)). This result indicates that $V_{frame}$ functions as an index capturing play quality (i.e., the level of offensive opportunity), and further suggests that, for gaining an advantage in a game, it may be more important to stably and frequently generate plays above a certain quality threshold (medium to high level) than to produce a small number of extremely high-quality plays.

By contrast, for $V_{timing}$, which represents the optimality of the actually chosen initiation timing (Figure~\ref{fig:wuppcf_results}(b)), no marked difference was observed between the two teams. In other words, the distribution of play quality arising from the observed timing choices alone is insufficient to fully account for the five-point difference in the final score, implying that additional factors such as throwing accuracy or receiver movements other than timing may have contributed to the formation of the score gap.

\subsubsection{Within-possession dynamics: High-wUPPCF area ratio vs. turnovers \& scoring}
\begin{figure}[b]
    \centering
    \includegraphics[width=1\linewidth]{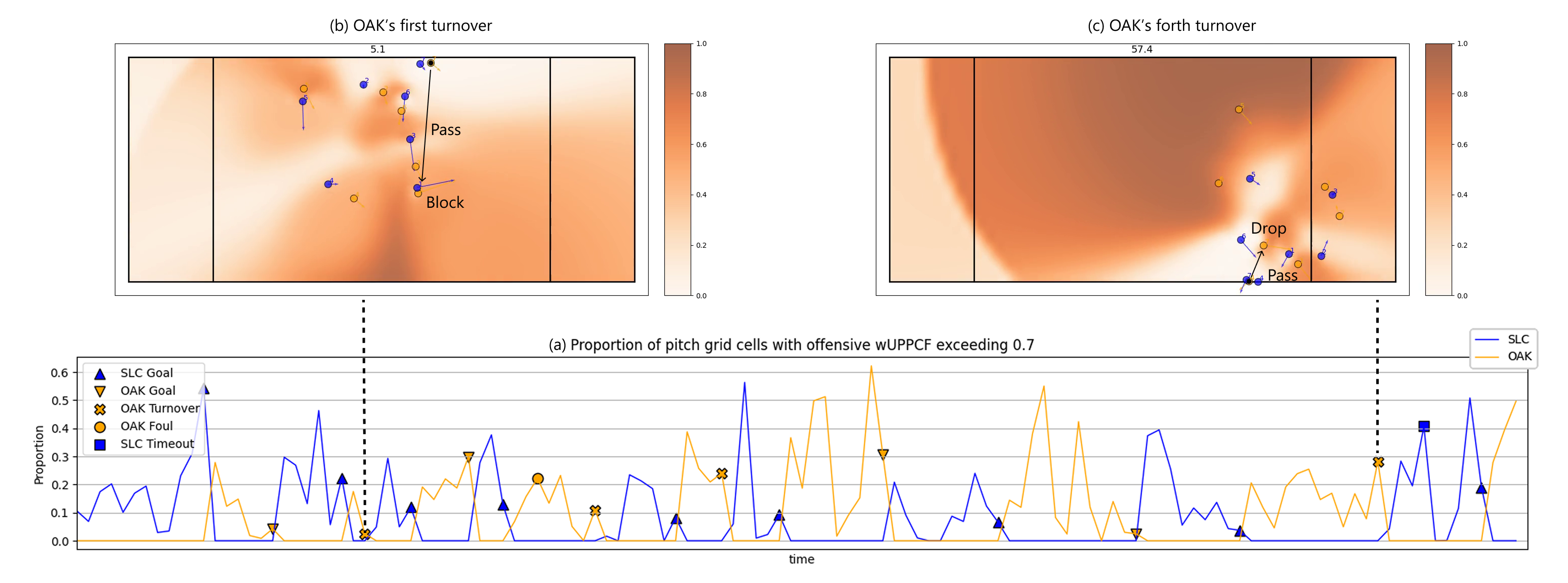}
    \caption{ (a) The line plot in the lower panel shows, for all possessions in the second quarter, the proportion of grid cells with wUPPCF values exceeding 0.7 (High-wUPPCF area ratio) for each team at 5-second intervals, visualizing how safe passing areas for the offense are created over time. Larger values indicate that a wider region of the field is under favorable pitch control for the offense.
    (b) Example of OAK’s first turnover. In a situation where the offense has limited advantageous space, a pass is thrown into an area with strong defensive pressure, resulting in the disc being blocked.
    (c) Example of OAK’s fourth turnover. Although a sufficient amount of advantageous space is available for the offense, the turnover is caused by an offensive error. As shown in (a), all six possessions in which the High-wUPPCF area ratio exceeded 0.4 at least once resulted in goals, whereas among the eight possessions in which it never exceeded 0.3, only five resulted in goals.}
    \label{fig:sequence}
\end{figure}

Figure~\ref{fig:sequence} illustrates, for all possessions in the second quarter, the proportion of grid cells with wUPPCF values exceeding 0.7 (hereafter, the “High-wUPPCF area ratio”) at 5-second intervals, thereby visualizing how the offense creates safe passing areas (advantageous space) over time. Larger values indicate that a wider region of the field is under favorable pitch control for the offense.

Focusing on OAK’s four turnovers, we observe that in three of them the High-wUPPCF area ratio remained at or below 0.3 throughout the possession, and in particular, the first and second turnovers were characterized by persistently low values over the entire possession. Both of these turnovers occurred when the defense made contact with the disc. In contrast, for the third and fourth turnovers, the High-wUPPCF area ratio during the possession did not drop as low; in these cases, the defense did not touch the disc, and the turnovers instead resulted from ``self-inflicted'' offensive mistakes such as throwing errors. Taken together, these observations suggest that the High-wUPPCF area ratio can serve as an indicator that characterizes situations in which, under strong defensive pressure, the offense fails to generate sufficiently safe passing areas.

With respect to scoring outcomes, all six possessions in which the High-wUPPCF area ratio exceeded 0.4 at some point eventually resulted in goals. In contrast, among the eight possessions in which the High-wUPPCF area ratio never exceeded 0.3, only five led to goals. Although the sample size is limited, this case study indicates that, at least in this match, possessions in which the offense attains a high High-wUPPCF area ratio and consistently secures wide safe passing areas tend to be more likely to result in scoring.

We also note that the High-wUPPCF area ratio sometimes temporarily decreases at the moment of scoring. This occurs because the evaluation at the end of the possession is based on the frame in which the disc is held inside the end zone after the goal has been scored, a state in which the offense is no longer attempting to create additional space. Therefore, such transient decreases in the HighwUPPCF area ratio immediately after the goal should be regarded as having no substantive implications for evaluating the offense’s space-creation ability.

\subsubsection{Case study: timing counterfactual on a single possession}
Here we present a UFA case study that shows how to read the outputs of our platform in a practical setting. Starting from synchronized video, tracking, and events, we examine the possession timeline, identify the receiver and the candidate initiation window, and summarize the resulting values. The aim is to demonstrate a careful interpretation of the metrics, including what changes when the start is earlier or later, which assumptions matter, and where the method is sensitive.

\begin{figure}[t]
    \centering
    \includegraphics[width=\linewidth]{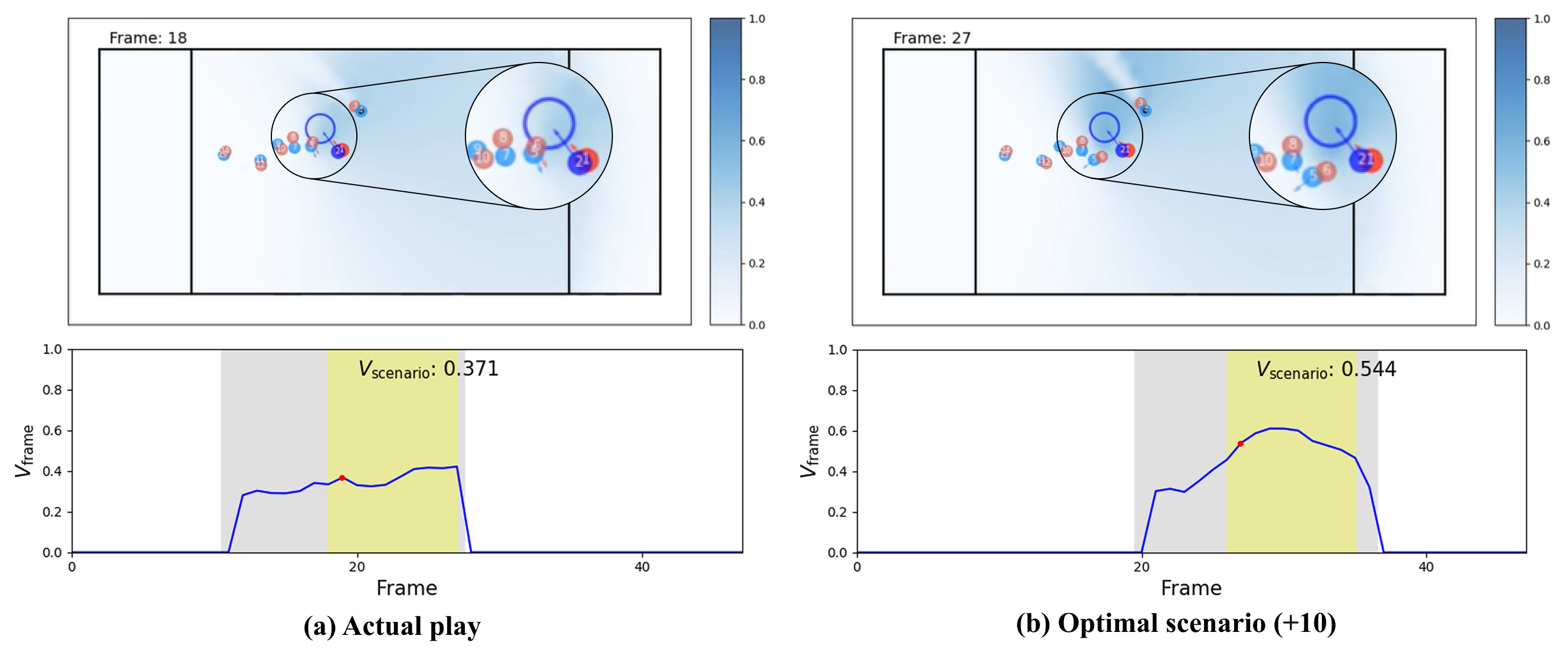}
    \caption{Off-ball initiation timing in UFA. (a) actual run; (b) counterfactual with the receiver’s start delayed by +10 frames. Blue markers indicate offensive players, red markers indicate defensive players, and the black marker denotes the disc. The darker-colored markers represent the offensedefense pair corresponding to the detected movement initiation. $V_{frame}$ is defined as the average wUPPCF value within the blue circle (with darker blue in the background indicating a more advantageous situation for the offense). The graph at the bottom shows the temporal evolution of $V_{frame}$, and the maximum of its 10-frame moving average is defined as $V_{scenario}$.}
    \label{fig:ultimate_application}
\end{figure}

Figure~\ref{fig:ultimate_application} presents an example of the movement-initiation analysis in Ultimate, comparing an actual play with an optimal scenario in which the receiver’s initiation is delayed by 10 frames (1 second).

$V_{frame}$ denotes the frame-wise space-evaluation score; at each frame, we quantify the value of the region controllable by the attacking player by integrating the positions and velocities of both teammates and opponents.  $V_{scenario}$ is defined as the maximum of the 10-frame moving average of $V_{frame}$, and is used as an overall evaluation index for a single cutting scenario.

In the actual play, the receiver initiated movement before the offensive and defensive players
ahead had fully cleared the lane (Figure~\ref{fig:ultimate_application}(a)), which restricted the amount of space that could be
effectively exploited. In contrast, in the optimal scenario, delaying the initiation by 10 frames
allowed the receiver to cut into the space vacated by those players, resulting in an increase in the
scenario evaluation value $V_{scenario}$ from 0.371 to 0.544 (Figure~\ref{fig:ultimate_application}(b)).

Thus, this framework, which explicitly manipulates and evaluates the receiver’s initiation timing, enables us to quantitatively indicate at what timing movement initiation maximizes spatial utilization, and suggests that it can provide interpretable and concrete feedback to coaches and players. More broadly, the case study illustrates how a unified, sport-agnostic pipeline can surface timing effects that are otherwise difficult to isolate from raw trajectories, demonstrating how the same analysis can be applied across sports without sport-specific rewrites. By showing how counterfactual timing shifts propagate through space valuations and scenario metrics, the example highlights the platform’s potential to support consistent diagnostic workflows and comparable interpretations in many sports.

\begin{table}[b]
\captionsetup{width=\textwidth,justification=raggedright,singlelinecheck=false}
\caption{The Pearson correlation between the current ($i$-th) and the subsequent ($i+1$-th) game for teams in the World Cup dataset for the four performance indicators: OBSO, BIMOS, Shots, and Goals.}
\label{tab:obso_bimos_transition}
\begin{tabular}{ccccc}
\toprule
$i\textbackslash i{+}1$ & OBSO & BIMOS & Shots & Goals \\
\midrule
OBSO  & 0.39 & 0.76  & 0.63  & 0.16  \\
BIMOS & -    & 0.48  & 0.82  & 0.18  \\
Shots & -    & -     & 0.55  & 0.20  \\
Goals & -    & -     & -     & 0.41  \\
\bottomrule
\end{tabular}
\end{table}

\subsection{Cross-Sport Transfer: OBSO vs. BIMOS in Soccer}
Next, we benchmarked BIMOS against OBSO on World Cup Soccer data to assess the transfer. The analysis focused on team-level temporal stability and predictive relationships, probability calibration, and a surface comparison that contrasts a static-target view (PPCF × Transition) with a dynamic delivery view (PBCF) to see where each model better captures the play.

\subsubsection{Temporal stability \& predictive relationships}
The Pearson correlation in Table~\ref{tab:obso_bimos_transition} provides insights into the validity of the OBSO and BIMOS models. The diagonal entries show that both OBSO ($0.39$) and BIMOS ($0.48$) exhibit moderate temporal stability across matches, comparable to goals ($0.41$) but lower than shots ($0.55$), indicating that these models capture repeatable team behaviors rather than match-specific noise. The cross-metric trends further highlight their predictive value, where both OBSO and BIMOS show substantial correlation with future shooting volume ($0.63$ and $0.82$), outperforming goals ($0.20$). These results suggest that these two models capture upstream attacking intent that preceded shot creation. As expected, correlations with future goals are relatively weak due to the stochastic nature of scoring. Taken together, OBSO and BIMOS are robust and informative indicators for sustainable attacking intent, offering more predictive insight than raw goal outcomes.

The log-loss results further show the differences in goal prediction accuracy between the two models. BIMOS achieves a substantially lower log-loss value ($0.3943$) compared to OBSO ($0.6508$), indicating that BIMOS produces a more consistent estimate. Because log-loss penalizes both overconfident errors and underconfident predictions, the lower value suggests that BIMOS generated better-calibrated scoring probabilities across matches. OBSO, while still informative for explaining attacking intent, produces less accurate probability estimates when evaluated directly against goal events.

\subsubsection{Surface comparison: PPCF × Transition vs. PBCF}
\begin{figure}[t]
    \centering
    \begin{subfigure}{0.48\textwidth}
        \captionsetup{justification=centering}
        \centering
        \includegraphics[width=\textwidth]{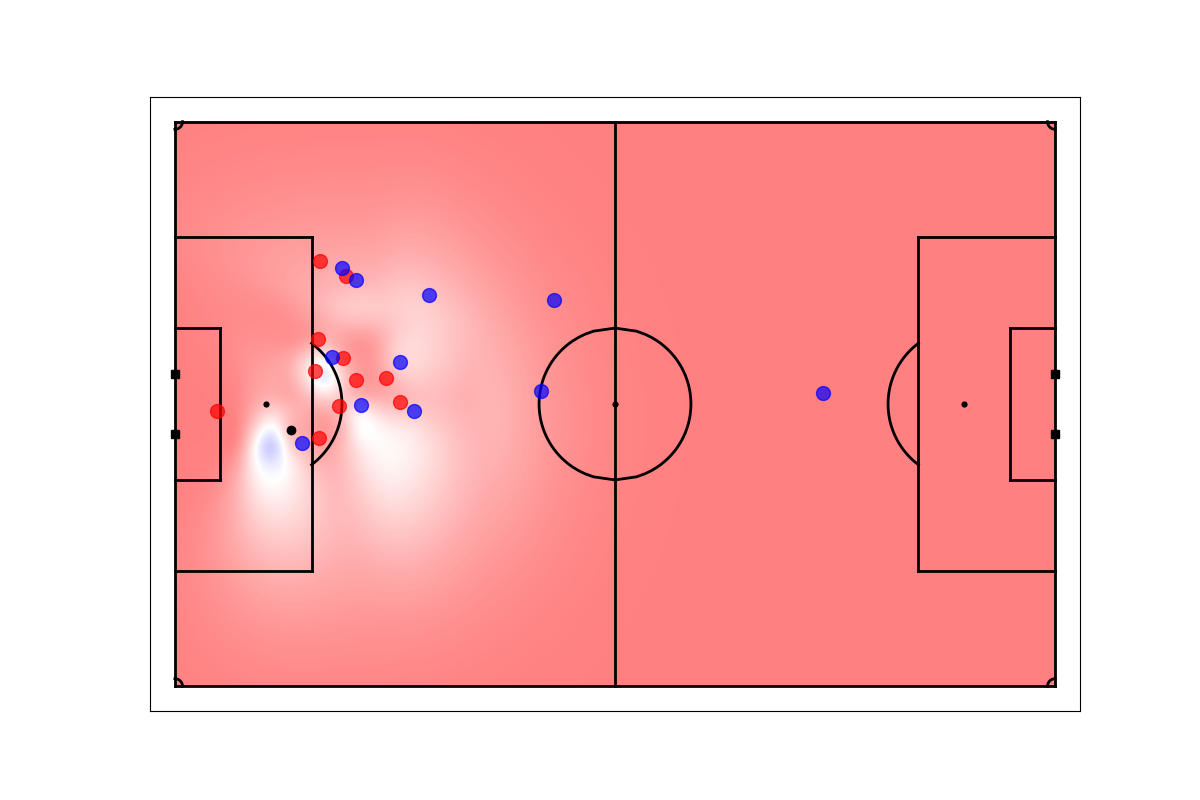}
        \caption{PPCF$\times$Transition}
        \label{fig:sub1}
    \end{subfigure}
    \hfill
    \begin{subfigure}{0.48\textwidth}
        \captionsetup{justification=centering}
        \centering
        \includegraphics[width=\textwidth]{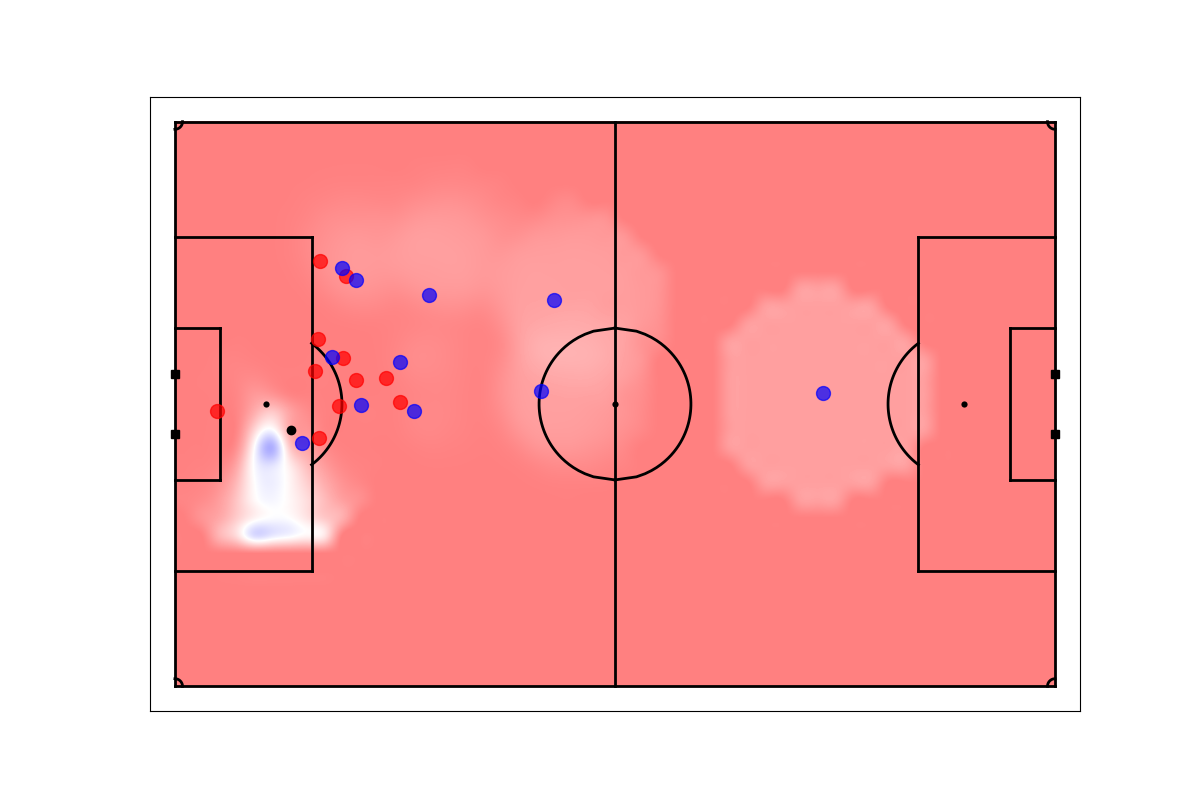}
        \caption{PBCF}
        \label{fig:sub2}
    \end{subfigure}

    \caption{Comparison of spatial control estimations using (a) PPCF multiplied by Transition probability in OBSO \citep{Spearman18} and (b) PBCF in BIMOS \citep{kono2024mathematical} for the same match (shot) situation. Blue and red dots represent attacking and defending players, respectively, while the colored surfaces illustrate the model-estimated control probabilities across the pitch. PBCF reveals defensive strengths along the actual ball trajectory, beyond the static target assumption in PPCF and transition model.}
    \label{fig:compare_ppcf_pbcf}
\end{figure}

Figure~\ref{fig:compare_ppcf_pbcf} compares two spatial control representations for the same match (shot) situation: PPCF multiplied by the Transition probability and PBCF. The PPCF × Transition map reflects regions where each team is more likely to gain control at a fixed target location under a lobbed-trajectory assumption. In contrast, PBCF updates the “ball location"continuously along the delivery path and evaluates, at each time horizon, which player can intercept the ball first rather than simply reach a target location. As defenders are frequently positioned between the passer and the intended direction of play, PBCF typically indicates stronger defensive intervention capability relative to PPCF × Transition. Consequently, PBCF remains broadly similar to PPCF × Transition but reveals more nuanced defensive strengths along the actual ball trajectory, where defenders can intervene earlier, particularly within congested areas in front of goal.

\section{Conclusion}
We introduced an open, pip-installable platform that standardizes professional tracking into a common format and delivers sport-agnostic spatial evaluations with sport-specific options, which enables reuse across Ultimate, basketball, and soccer. Within this platform, OBSO, CRSV, and BIMOS provide complementary views of chance creation and usable space.
In Ultimate, team-level distributions and within-possession dynamics were consistent with the space measures, and a single-possession counterfactual illustrated how varying initiation timing changes the evaluation.
In soccer, adapting a basketball model (BIMOS) produced probability estimates more consistent with observed outcomes than OBSO, indicating improved calibration under cross-sport transfer.
Taken together, these results support a unified, timing-aware approach to spatial evaluation that travels across sports and yields interpretable outputs for analysis. Future work will expand datasets and sports, incorporate individual-specific motion and reaction parameters, and examine real-time and analyst-in-the-loop use.

\backmatter

\bibliography{main}

\end{document}